\documentclass[fleqn,usenatbib]{mnras}

\pdfoutput=1
\usepackage{mathptmx}
\usepackage[T1]{fontenc}
\usepackage{ae,aecompl}

\usepackage{graphicx}	% Including figure files
\usepackage{amsmath}	% Advanced maths commands
\usepackage{amssymb}	% Extra maths symbols
\usepackage{multirow}
\usepackage{hyperref}
\hypersetup{colorlinks=true,linkcolor=blue,citecolor=blue,filecolor=blue,urlcolor=blue}

\newcommand{\zxsun}{(Z/X)_\odot}
\newcommand{\lsun}{L_\odot}
\newcommand{\rsun}{R_\odot}

\title[Solar models from solar wind abundances]{Implications of solar wind measurements for solar models and composition}

\author[A. Serenelli et al.]{Aldo Serenelli,$^{1}$\thanks{E-mail: aldos@ice.csic.es}
Pat Scott,$^{2}$
Francesco L. Villante, $^{3,4}$
Aaron C. Vincent,$^{5}$ \newauthor
Martin Asplund,$^{6}$
Sarbani Basu,$^{7}$
Nicolas Grevesse,$^{8,9}$
and Carlos Pe\~na-Garay,$^{10,11}$
\\
$^{1}$Institute of Space Sciences (IEEC-CSIC), Carrer de Can Magrans, Barcelona, E-08193, Spain \\
$^{2}$Department of Physics, Imperial College London, Blackett Laboratory, Prince Consort Road, London SW7 2AZ, UK \\
$^{3}$Dipartimento di Scienze Fisiche e Chimiche, Universit\`a dell'Aquila, Via Vetoio, I-67100 L'Aquila, Italy \\
$^{4}$Istituto Nazionale di Fisica Nucleare (INFN), Laboratori Nazionali del Gran Sasso (LNGS), Via G. Acitelli, I-67100 Assergi (AQ), Italy \\
$^{5}$Institute for Particle Physics Phenomenology (IPPP), Department of Physics, Durham University, Durham DH1 3LE, UK \\
$^{6}$Research School of Astronomy and Astrophysics, Australian National University, Canberra, ACT 2611, Australia \\
$^{7}$Department of Astronomy, Yale University, PO Box 208101, New Haven, CT 06520, USA \\
$^{8}$Centre Spatial de Li\`ege, Universit\'e de Li\`ege, avenue Pr\'e Aily, B-4031, Angleur-Li\`ege, Belgium \\
$^{9}$Institut d'Astrophysique et 
de G\'eophysique, Universit\'e de Li\`ege, All\'ee du 6 ao\^ut, 17, B5C, B-4000, Li\`ege, Belgium \\
$^{10}$Instituto de Fisica Corpuscular, CSIC-UVEG, P.O.  22085, Valencia, 46071, Spain \\
$^{11}$Laboratorio Subterr\'aneo de Canfranc, Estaci\'on de Canfranc, 22880, Spain
}

\date{Accepted XXX. Received YYY; in original form ZZZ}

\pubyear{2015}

\begin{document}
\label{firstpage}
\pagerange{\pageref{firstpage}--\pageref{lastpage}}
\maketitle

% Abstract of the paper
\begin{abstract}
We critically examine recent claims of a high solar metallicity by \citet[vSZ16]{vS16} based on \textit{in situ} measurements of the solar wind, rather than the standard spectroscopically-inferred abundances \citep[AGSS09]{agss09}.  We test the claim by \citet{v16} that a composition based on the solar wind enables one to construct a standard solar model in agreement with helioseismological observations and thus solve the decades-old solar modelling problem. We show that, although some helioseismological observables are improved compared to models computed with spectroscopic abundances, most are in fact worse. The high abundance of refractory elements leads to an overproduction of neutrinos, with a predicted $^8$B flux that is nearly twice its observed value, and $^7$Be and CNO fluxes that are experimentally ruled out at high confidence. A combined likelihood analysis shows that models using the vSZ16 abundances fare worse than AGSS09 despite a higher metallicity. We also present astrophysical and spectroscopic arguments showing the vSZ16 composition to be an implausible representation of the solar interior, identifying the first ionisation potential effect in the outer solar atmosphere and wind as the likely culprit. 
\end{abstract}

% Select between one and six entries from the list of approved keywords.
% Don't make up new ones.

%%%%%%%%%%%%%%%%%%%%%%%%%%%%%%%%%%%%%%%%%%%%%%%%%%

%%%%%%%%%%%%%%%%% BODY OF PAPER %%%%%%%%%%%%%%%%%%

\section{Introduction}

Solar model atmospheres based on three-dimensional (3D) radiation hydrodynamic simulations of near-surface convection have been central in the determination of element abundances in the solar photosphere for the last 15 years (see e.g. reviews by \citealp{2005ARA&A..43..481A, 2009LRSP....6....2N}). Further work in the field has included the introduction of non-local thermodynamic effects (non-LTE) in the modelling of atomic line formation. Together with improvements in atomic physics, these efforts have led to a reduction in the abundance of elements determined from the solar photospheric spectrum. CNO elements are particularly affected and recommended values \citep[AGSS09]{agss09} are 40\% lower than previous determinations \citep[GS98]{gs98}. Abundances of refractories, e.g. Mg, Si, Fe, have also been reduced, by smaller amounts, typically around 10\% \citep{Scott15a,Scott15b,Grevesse15}. 

Standard solar models (SSMs) use the present-day solar photospheric composition, in the form of the metal-to-hydrogen abundance ratio $\zxsun$, as one of the constraints that must be fulfilled by construction. The other two constraints are the solar luminosity ($\lsun$) and radius ($\rsun$). The SSM is the result of the evolution of a 1~M$_\odot$ stellar model from the pre-main sequence to the solar system age $\tau_\odot$. In order to match $\zxsun$, $\lsun$ and $\rsun$, the initial helium ($Y_{\rm ini}$) and metal ($Z_{\rm ini}$) mass fractions and the mixing length parameter ($\alpha_{\rm MLT}$) are adjusted in the model. SSMs account for ``standard'' physics in stellar models, and avoid inclusion of physics that needs phenomenological calibration as much as possible; convection is the most important exception. SSMs used in this work are based on the same input physics as those in \citet{serenelli:2011}, except for the choice of solar composition.

$\lsun$ and $\rsun$ are well known, with very small uncertainties. The constraint on $\zxsun$, which is less well measured, thus determines both the metallicity as well as the Sun's helium abundance; solar models constrained to fulfill GS98 (high-$Z$) or AGSS09 (low-$Z$) compositions show quite different internal structures. In particular, solar models based on the AGSS09 composition show strong discrepancies with helioseismic probes of the solar interior, whereas the older GS98 composition leads to quite nice agreement \citep[among many]{serenelli:2009}. This discrepancy is known as the ``solar modelling problem''. The seismic probes relevant to the solar modelling problem include: the profile of the solar sound speed, the depth of the convective envelope, the surface helium abundance and frequency separation ratios. 

This solar modelling problem has been around for more than 10 years and, while there is no lack of proposed ideas, no definitive solution has yet been found. 
It is well known that the comprehension of the solar abundance problem is intimately related to understanding 
the role of opacity in solar modelling, since the effects produced by a modification of the radiative opacity are almost degenerate (with the notable
exception of CNO neutrinos) with those produced by a modification of the heavy element admixture. 
Namely, the agreement with helioseismology using the AGSS09 composition could be restored by a suitable modification 
of the opacity profile of the Sun, as described in \citet{villante:2010b} and \citet{villante:2014}.
The current generation of standard solar models typically relies on opacities from the Opacity Project \citep[OP, e.g.][]{2005MNRAS.360..458B} or OPAL \citep[e.g.][]{1996ApJ...464..943I}.  Recently \citet{2015Natur.517...56B} experimentally measured the opacity at conditions similar to those immediately below the solar convection zone for the first time, finding wavelength-dependent Fe opacities typically 30--40\% higher than predicted by OP and OPAL. When folded into the Rosseland mean opacity needed in solar model calculations, the result is a $7\pm3\%$ increase. Such extra opacity goes towards solving the solar modelling problem by itself. Recently, also, theoretical work by \citet{krief:2016} found that line broadening induces changes in the opacity profile in a solar model that mimics the opacity variations required by helioseismic constraints \citet{villante:2014}. Detailed solar modelling however will have to await a better understanding of how such opacity increases depend on the physical conditions (temperature and density) for the relevant elements.

\citet[hereafter vSZ16]{vS16} have presented results from in-situ measurements of the chemical composition of the solar wind. In particular, they have determined abundances of C, N and O, as well as the most abundant refractories: Mg, Si, S and Fe. The comparison between elemental abundances from GS98, AGSS09 and vSZ16 is presented in Table~\ref{tab:compo}. 
\citet{v16} (V16 hereafter) analyse the impacts of the vSZ16 abundances on solar interior modelling. They focus in particular on a certain subset of helioseismic probes, but with special emphasis on the solar sound speed profile. Their analysis is based on the so-called linear solar models, originally developed by \citet{villante:2010}. V16 reach the conclusion that, when the vSZ16 composition is used in solar models, the agreement between solar models and helioseismic probes is restored, and claim that they have found a solution to the solar abundance problem. Here we present a number of arguments against this claim. In fact, the vSZ16 composition, when used to construct an SSM as proposed by V16, produces a model that is strongly at odds with experimental evidence, actually leading to an even larger discrepancy than models based on AGSS09 abundances.  We quantify the disagreement both with neutrino and helioseismological observations, and then go on to outline a number of other reasons why the vSZ16 abundances cannot be representative of the actual solar composition.

We note that \cite{caffau:2011} have also published solar photospheric abundances based on 3D model atmospheres. For the two most important volatiles, C and O, their results are 0.07 dex larger than the values of AGSS09. However, we do not refer to those results further here for the following reasons. The first one is that \cite{caffau:2011} does not provide a complete determination of the solar mixture and that implies using an alternative source for the missing elements. Also, among the missing elements is Si, the anchor point between the photospheric and meteoritic scales, our preferred choice building SSMs\footnote{Helioseismic and solar neutrino results of a SSM built with a hybrid solar mixture based on \cite{caffau:2011} abundances, however, has been recently discussed in \cite{serenelli:2016}.}. Finally, articles V16 and vSZ16 do not use these data in their core results and our main aim in this work is to evaluate how SSMs constructed with the vSZ16 composition compare to helioseismic and solar neutrino data, not to discuss a spectrum of possible (partial) solutions to the solar modelling problem.

The rest of this paper is organised as follows. Section~\ref{sec:neutrinos} discusses in some detail the implications of vSZ16 abundances on SSM predictions for neutrino fluxes.
Solar neutrino fluxes were not considered by V16 except for a rough estimate, which significantly underestimates the neutrino fluxes predicted by their solar model. Solar neutrinos are very important probes of the solar interior, and in particular of the core, where vSZ16 abundances strongly affect solar model properties. In Section~\ref{sec:helios}, we discuss the helioseismic properties of a standard solar model with the vSZ16 composition. Whereas V16 focused on the sound speed profile and, to a lesser extent, on the surface helium abundance and depth of the convective zone, we extend our analysis to ratios of separation frequencies that are well-known probes of the solar core \citep{basu:2007,chaplin:2007}.  Section~\ref{sec:combined} presents a combined likelihood analysis of solar models with different candidate compositions.  Section~\ref{sec:compo} can be read independently of those related to the SSM; here we present a number of arguments that show that it is in fact very unlikely that vSZ16 abundances are representative of the photospheric and interior composition of the Sun. We end with a summary of our most relevant findings, all of which point to the same conclusions: that vSZ16 is not representative of the photosphere, and that SSMs based on vSZ16 are in disagreement with both helioseismic probes of the solar interior and solar neutrino fluxes.  Despite a metallicity that is closer to the classic GS98 measurement, they therefore perform worse than AGSS09.

\begin{table}
\centering
\caption{Adopted solar chemical compositions. Abundances given as $\log{\epsilon_i}\equiv\log{N_i/N_H}+12$.  AGSS09ph refers to the solar photospheric abundances from AGSS09, whereas AGSS09met refers to the case when the abundances for all non-volatile elements (i.e. everything other than C, N, O and the noble gases) are replaced with the corresponding CI carbonaceous chondrite meteoritic abundances from \citet{Lodders09}.  Note that the errors attached to vSZ16 values incorporate the error on the absolute scale, but do not incorporate additional systematics expected from unquantified fractionation effects (see Sec.\ \protect\ref{abundances}). \label{tab:compo}}
\begin{tabular}{lcccc}
\hline 
 & \multicolumn{4}{c}{$\log{\epsilon}$} \\ 
\cline{2-5}
Element & GS98 & AGSS09met & AGSS09ph & vSZ16 \\ \hline
C & 8.52  & $8.43\pm 0.05 $& $8.43\pm0.05$& $8.65^{+0.11}_{-0.09}$\vspace{0.5mm}\\
N & 7.92  & $7.83\pm 0.05 $& $7.83\pm0.05$& $7.97^{+0.15}_{-0.11}$\vspace{0.5mm}\\
O & 8.83  & $8.69\pm 0.05 $& $8.69\pm0.05$& $8.82^{+0.10}_{-0.08}$\vspace{0.5mm}\\
Ne & 8.08 & $7.93\pm 0.10 $& $7.93\pm0.10$& $7.79^{+0.17}_{-0.12}$\vspace{0.5mm}\\
Mg & 7.58 & $7.53\pm 0.01 $& $7.60\pm0.04$& $7.85^{+0.17}_{-0.13}$\vspace{0.5mm}\\
Si & 7.56 & $7.51\pm 0.01 $& $7.51\pm0.03$& $7.82^{+0.18}_{-0.13}$\vspace{0.5mm}\\
S & 7.20  & $7.15\pm 0.02 $& $7.12\pm0.03$& $7.56^{+0.19}_{-0.13}$\vspace{0.5mm}\\ 
Fe & 7.50 & $7.45\pm 0.01 $& $7.50\pm0.04$& $7.73^{+0.17}_{-0.12}$\vspace{0.5mm}\\
\hline
Z/X & 0.0229 & 0.0178 & 0.0181 & 0.0265 \\ \hline
\end{tabular}
\end{table}

\section{Solar neutrinos} \label{sec:neutrinos}

Solar neutrino fluxes predicted by solar models are particularly sensitive to variations in the solar composition. For pp-chain fluxes this dependence occurs through the impact of metals on the radiative opacity profile of the Sun. Therefore, individual elements will have different impacts on neutrino predictions. Scaling $\zxsun$ globally is not a correct way to estimate the variations induced in solar neutrinos when the composition is varied. This was discussed at length by \citet{bahcall:2005}; since then the detailed composition has been used in studies of solar neutrinos.

Before discussing results based on SSMs, we present simple but quite accurate estimates of the changes in the neutrino fluxes expected when changing from AGSS09 to vSZ16 abundances, based on power-law expansions \citep{bahcall:1989}. As an example, let us consider the flux of neutrinos from $^8{\rm B}$ decay in the pp chain, $\Phi(^8{\rm B})$, and focus only on the most relevant elements. The dependence of the $\Phi(^8{\rm B})$ flux on variations of elemental abundances is given by the following power-law exponents: 0.139, 0.109, 0.092, 0.192, 0.140, 0.502 for O, Ne, Mg, Si, S and Fe respectively \citep{serenelli:2013}. Taking AGSS09ph as the reference composition, the expected fractional increase in $\Phi(^8{\rm B})$, if the vSZ16 composition is used instead, is
\begin{align}
\frac{\delta \Phi{(^8{\rm B})}}{\Phi{(^8{\rm B})}} \approx 0.139\times0.35 + 0.109\times(-0.28) + 0.092 \times 0.78  \nonumber \\
\phantom{0} + 0.192 \times 1.04 + 0.140 \times 1.75 + 0.502 \times 0.70 = 0.89. \nonumber
\end{align}
Here, we have taken the fractional variations for the composition directly from Table~1 in V16.  The same comparison against the AGSS09met composition yields an even larger variation: $\delta \Phi(^8{\rm B})/\Phi(^8{\rm B}) = 0.99$. The power-law expansion thus leads to an estimated $90-100\%$ increase in $\Phi(^8{\rm B})$ for an SSM based on vSZ16. 

Power-law expansions describe the first order response of solar models to changes in the input parameters. The changes between AGSS09 and vSZ16 are not small, and it might be that second order effects play a significant role. In the rest of this article, we therefore present results based solely on full SSMs that consistently account for the adopted solar composition. We have calibrated an SSM using the vSZ16 abundances and the same input physics as described in \citet{serenelli:2011}. In addition, we also use the two SSMs computed in that work with the GS98 and the AGSS09 (more precisely AGSS09met, see Table \ref{tab:compo}) compositions.

Results for the three SSMs and all neutrino fluxes are summarised in Table~\ref{tab:neutrinos}. Solar values for the neutrinos from all pp-chains are those recently determined by \citet{bergstrom:2016} and result from a combined analysis using all possible sources of experimental data. For $^{13}$N, $^{15}$O and $^{17}$F fluxes we quote the upper 68\% limit.  As can be seen in Table\,\ref{tab:neutrinos}, the SSM based on vSZ16 predicts $\Phi(^8{\rm B}) = {\rm (10.1 \pm 1.8) \times10^6\,cm^{-2}\,s^{-1}}$, similar to but slightly larger than the simple power-law estimation. This is more than a factor of 2 increase with respect to the AGSS09 model and much higher than the 10\% increase guessed, with no further justification, in V16.

\begin{table}
\centering
\caption{Solar neutrinos fluxes in ${\rm cm^{-2}\,s^{-1}}$ and relative errors derived from some standard solar models (with different compositions) 
and from oscillation neutrino data; Units are: $10^{10}$ (pp), 
$10^{9}$ ($^{7}$Be), $10^{8}$ (pep, $^{13}$N, $^{15}$O), $10^{6}$ ($^{8}B$, $^{17}$F) and $10^{3}$ (hep).\label{tab:neutrinos}} 
\begin{tabular}{lcccc}
\hline
Source & GS98, AGSS09met & vSZ16 & Solar \\ \hline
pp & $5.98,\,6.03(1\pm0.006)$ & $5.78 (1\pm0.008)$& $5.97(1\pm0.005)$ \\
pep & $1.44,\,1.47(1\pm0.012)$ &  $1.34 (1\pm0.016)$& $1.45(1\pm 0.009)$ \\
hep & $8.04,\,8,31(1\pm0.30)$ & $7.23(1\pm0.30)$ & $19(1\pm0.55)$ \\
$^7$Be & $5.00,\,4.56(1\pm0.07)$ & $6.58 (1\pm0.08)$& $4.80(1\pm 0.05)$ \\
$^8$B & $5.58,\,4.59(1\pm0.14)$  & $10.1 (1\pm0.18)$& $5.16(1\pm 0.022)$ \\
$^{13}$N & $2.96,\,2.17 (1\pm0.14)$ &  $5.46  (1\pm0.21)$&  $\leq$\,13.7 \\
$^{15}$O & $2.23,\,1.56 (1\pm0.15)$ & $4.56  (1\pm0.22)$&  $\leq$\,2.8\\
$^{17}$F & $5.52,\,3.40 (1\pm0.17)$ & $9.01  (1\pm0.30)$&   $\leq$\,85 \\ \hline
\end{tabular}
\end{table}

The large change in $\Phi(^8{\rm B})$ is due to the large fractional increase in abundances of refractories proposed by vSZ16. 
Refractories such as Mg, Si, S, Fe have relatively high atomic charge and are therefore important contributors to the radiative opacity, even at the larger temperatures present in the solar core \citep{basu:2008,villante:2014}. The temperature in the core is therefore strongly correlated with the abundance of refractories.  Neutrino fluxes, especially those that depend very strongly on temperature, show an even more intense dependence.

The neutrino fluxes predicted by the vSZ16 solar model strongly disagree with the experimental constraints, 
as can be seen in Table\,\ref{tab:neutrinos}.
This statement can be quantified by returning to the example of  $\Phi(^8{\rm B})$. The observed flux is $\sim 50\%$ lower than the theoretical prediction. 
The uncertainty in the model flux not attributable to composition is of the order of 12\% \citep{serenelli:2013}, whereas the total uncertainty including errors on abundances is 14\% if AGSS09 uncertainties are adopted. The observed solar neutrino flux has an uncertainty of just 2\%. 
By combining experimental and theoretical uncertainties in quadrature,  we see that the disagreement is at $\sim 4\sigma$ level. This implies that there is simply no room in SSMs to bring the vSZ16 solar model prediction into agreement with the solar $\Phi{(^8{\rm B})}$. 
Only a strong reduction in refractories can bring down $\Phi{(^8{\rm B})}$, i.e. abandoning the vSZ16 composition. This reduction must bring refractories back to GS98 or AGSS09-like levels.

A similar reasoning follows for $\Phi(^7{\rm Be})$, for which the vSZ16 solar model predicts $6.58\times10^9\,{\rm cm^{-2}\,s^{-1}}$. This is again much higher than the experimental result, by about 4$\sigma$ when combining errors from modelling and experiment. Here again, the only way to bring this into agreement with the solar flux is to strongly reduce the abundance of refractories in the vSZ16 model to a level comparable to GS98 or AGSS09 (which agree quite well for non-volatile elements). 

There is an additional piece of information, relating to CN fluxes. Borexino \citep{bellini:2012} have presented the most restrictive limit to date on the sum of the $^{13}$N and $^{15}$O fluxes: $7.7\times10^8\,{\rm cm^{-2}\,s^{-1}}$, at 95\%~C.L. The sum of the $^{13}$N and $^{15}$O fluxes in the vSZ16 solar model is $10.02\times10^8\,{\rm cm^{-2}\,s^{-1}}$, i.e.\ well over the limit given by Borexino. This primarily occurs not because CN abundances are too high in the vSZ16 composition, but because refractories again push the model towards higher core temperatures, to which the CN fluxes are highly sensitive.

Here, we have compared predictions of SSMs to data. In \citet{villante:2014}, solar neutrino fluxes were used to infer the optimal solar composition by grouping elements together either as volatiles or refractories. All elements in each group were then  scaled by the same factor. These two factors were allowed to vary in a constrained manner so that both helioseismic models and solar neutrino data were reproduced. Results in that work, particularly the top right panel of Figure~6, clearly show that the typical 0.3\,dex increase in the abundance of refractories claimed by vSZ16, and used by V16 in analysing solar model predictions, is ruled out at a 6$\sigma$ level approximately. 

Furthermore, the direct comparison of solar neutrino fluxes with SSM predictions can be easily performed with the results of Table~\ref{tab:neutrinos}, adding errors in quadrature and using the experimental and theoretical error correlation matrix (see \citealp{serenelli:2011,bergstrom:2016} and references therein). The comparison of neutrino fluxes from model and data excludes the SSM with V16 abundances at more than 5$\sigma$, while 
neutrino data can not statistically differentiate the SSM with GS98 and AGSS09met abundances.

\section{Helioseismology} \label{sec:helios}

Next, we consider helioseismic tests of solar models. V16 employ linear solar models to study the response of the sound speed profile, depth of the convective zone and surface helium abundance to changes in the solar composition. We continue to base our results on full SSMs. It has to be made clear that the incorrect conclusions reached by V16 are not due to their use of linear solar models. However, an advantage of our approach here is that we can consider additional helioseismic probes that were not considered when linear solar models were originally developed, and were not taken into account in V16. 

\subsection{Sound speed profile.}

\begin{figure}
\includegraphics[scale=.275]{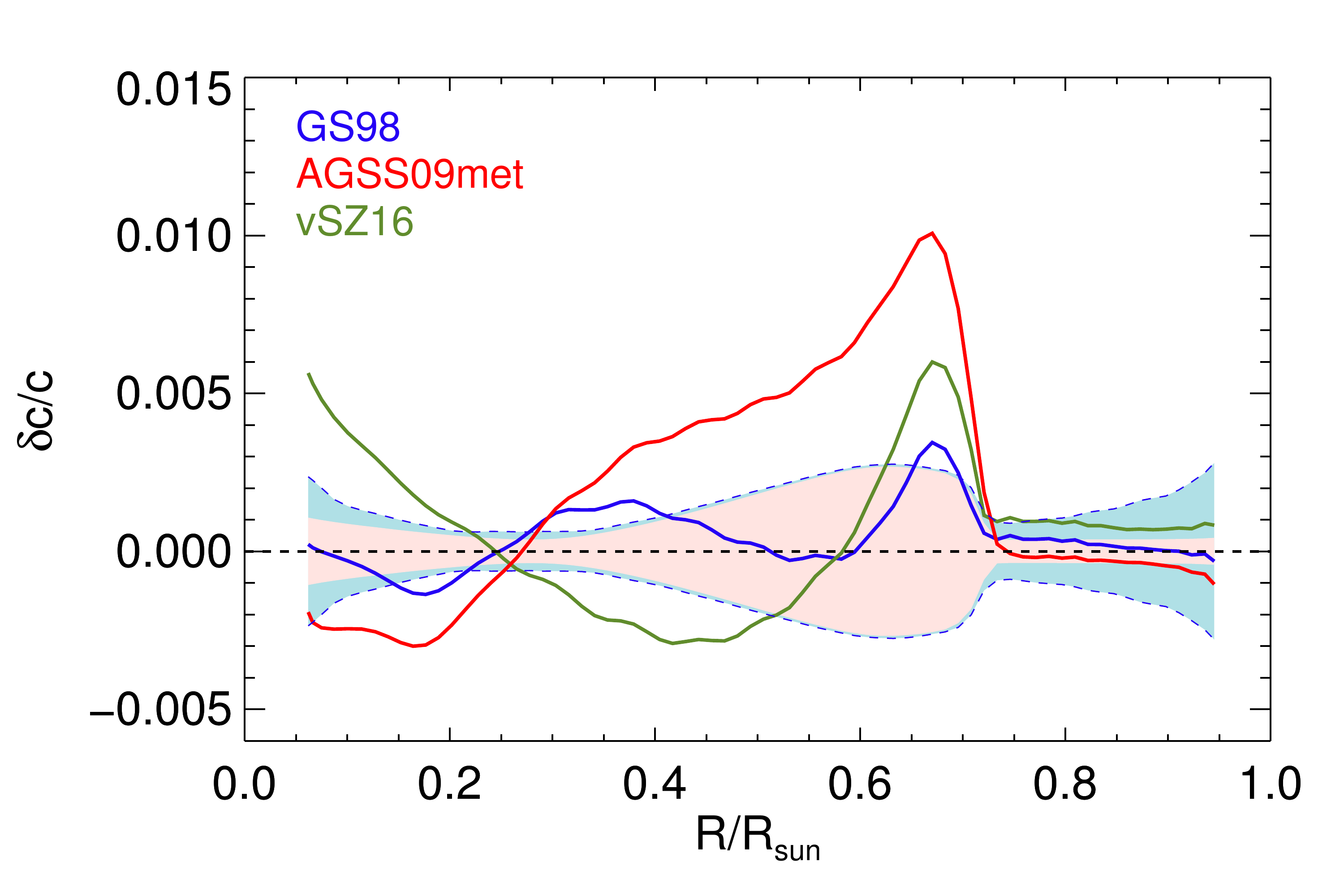}
\caption{The difference between the sound speed predicted by solar models and that obtained by inverting helioseismic observations, for models computed with three different compositions. The pink band shows the $1\sigma$ uncertainty due to the solar composition, whereas the blue band shows the total $1\sigma$ uncertainty.\label{fig:sound}}
\end{figure}

In Figure\,\ref{fig:sound} we plot the sound speed profiles for different solar models. The pink band shows 1$\sigma$ uncertainties associated with AGSS09 abundances. The light blue band indicates the combined uncertainty in solar models, statistical uncertainty (from uncertainties in solar frequency measurements) and systematic errors in the inversion procedure \citep{innocenti:1997}. For the latter, we have assumed what the authors refer to as the ``statistical'' approach, where different sources of uncertainty are assumed to be independent. V16 also considered this choice of errors, as well as the most conservative one proposed by \citet{innocenti:1997}. The specific choice of uncertainties is not, however, central to the arguments that follow. 

Results for the sound speed shown here are very close to those in V16. The agreement with helioseismic results is not nearly as good as for the GS98 composition, but it is clearly an improvement over AGSS09 at intermediate radii. This is not surprising, because the CNO abundances, O in particular, claimed by vSZ16 are quite close to GS98.  This is the main reason that the largest discrepancy between the AGSS09 model and the observed sound speed at around 0.7\,$\rsun$ is reduced when using the vSZ16 abundances.

Closer to the centre, the sound speed profile in the vSZ16 model starts deviating from the observed speed by a noticeable margin. The sound-speed estimates here are  indeed more uncertain, but current estimates of uncertainties are much less than those stated by Degl'Innocenti et al. (1997). For one, estimates of the frequencies of low-degree modes that penetrate the core are much better now than 20 years ago resulting in more precise inversion results \citep[see][]{basu:2009}. Additionally, a reduction of the error estimate, particularly in the convection zone, is a result of the realisation that inversion parameters need to be selected so as to minimise correlated errors between solutions at different radii.  This ensures that the solution is not biased, resulting in systematic errors (see \citealt{1996MNRAS.281.1385H,1999MNRAS.309...35R}). Also, there are now other probes of the structure of the innermost solar core that do not rely on inversion methods. These are the frequency separation ratios, which we turn to next, and which allow us to infer that inversion uncertainties from \citet{innocenti:1997} are likely to be overestimations of the actual errors.

\subsection{Small frequency separation ratios}

\cite{rox:2003} have shown that specific combinations of frequencies of low-degree modes can be used to construct helioseismic diagnostics that are largely insensitive to the structure of the outer layers of the Sun.  These are particularly sensitive to the structure of the innermost 10--15\% of the solar radius \citep[see the appendix of][]{basu:2007}. This is the region where sound speed inversions become more uncertain, and where the majority of the solar neutrinos are produced. These so-called small frequency separation ratios are given by
\begin{align}
r_{02}(n) = \frac{\nu_{n,0} - \nu_{n-1,2}}{\nu_{n,1}-\nu_{n-1,1}}; \ \ \ 
r_{13}(n) = \frac{\nu_{n,1} - \nu_{n-1,3}}{\nu_{n+1,0}-\nu_{n,0}},  
\end{align}
where $n$ represents the radial order of a mode and the second index is its angular degree.
 
\begin{figure*}
\includegraphics[scale=.28]{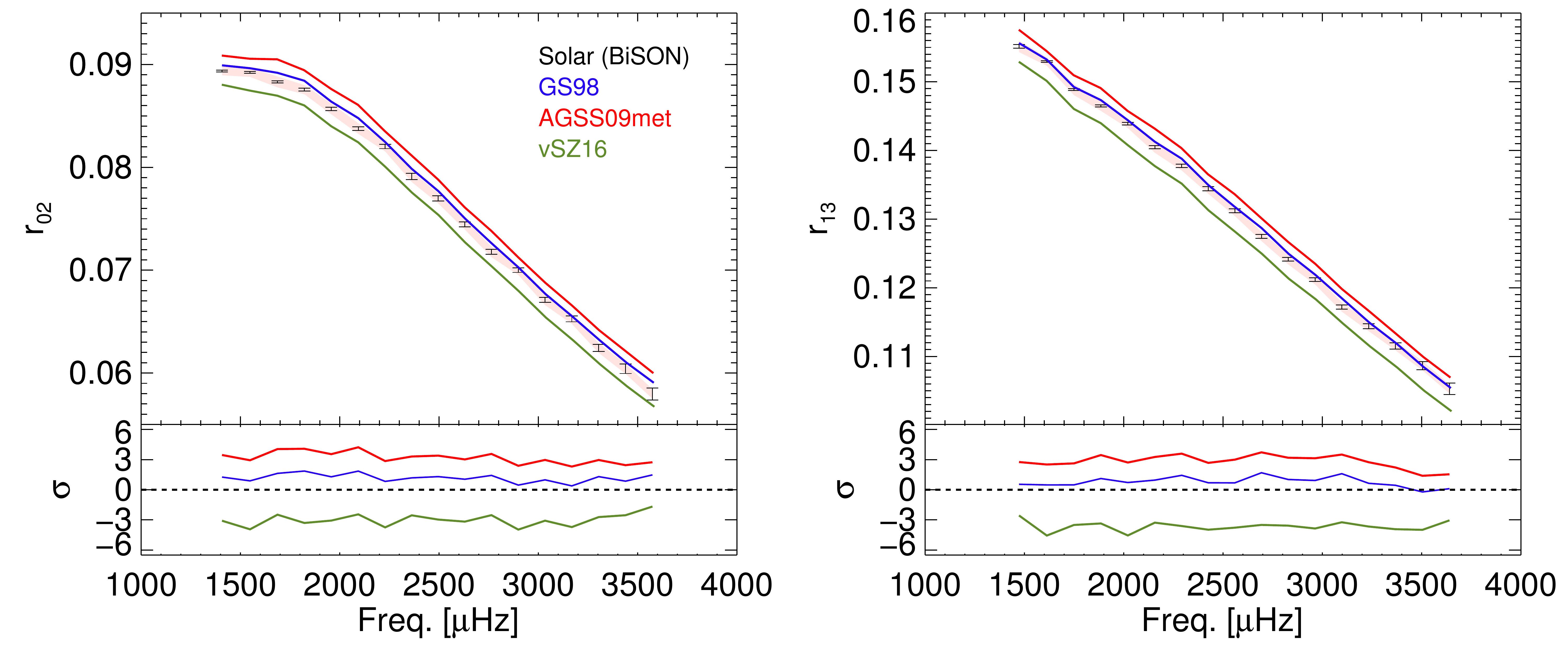}
\caption{Small frequency separation ratios for three SSMs. Solar data is depicted by points and error bars. The shaded band denotes 1$\sigma$ composition uncertainties associated with solar models, based on the uncertainties of the AGSS09met abundances. Lower panels show residuals, in units of the total combined (model and data) uncertainty for each frequency separation ratio. \label{fig:ratios}}
\end{figure*}

As mentioned above, the large increase in refractories proposed by vSZ16 has the strongest impact on the central regions of the Sun. This is hinted at by the shape of the sound speed profile (Fig.~\ref{fig:sound}), and  becomes obvious in the separation ratios which are shown in Figure~\ref{fig:ratios}, where results for the SSMs are shown with lines and those for the Sun \citep[based on 4752 days of BiSON data;][]{chaplin:2007,basu:2009} are shown with points and error bars. 
The pink shaded area denotes the 1$\sigma$ uncertainties in solar models, based on the solar abundance uncertainties as given by AGSS09. The discrepancy between the vSZ16 solar model and helioseismic data is clearly visible. In fact, it is does a slightly worse job overall than the AGSS09met solar model. The error band includes model errors from both the solar composition and other sources. Using the same model errors for the three SSMs, and assuming for simplicity that errors are uncorrelated, the combined $\chi^2$ of the 34 available points for $r_{02}$ and $r_{13}$ are: 41, 321, and 389 for the GS98, AGSS09met, and vSZ16 solar models. If, instead, only non-composition uncertainties in the models are taken into account, $\chi^2= \{76, 603, 705\}$, in the same ordering. The assumption of uncorrelated errors is of course not correct, but a proper account of correlations, that will be the same for all the models, will not affect the qualitative picture: vSZ16 not only fails at alleviating the solar modelling problem present for AGSS09met, but in fact performs worse. 

The conclusion is therefore the same as before: changes in the solar structure induced by a large increase in refractories cannot be compensated by any other variations in the SSM inputs. Only large changes in the composition can bring the vSZ16 model into agreement with data, such that refractories have to be close to those in GS98 and AGSS09.

As a final point, we consider the separation ratios and the sound speed profiles together. The GS98 model predicts both sound speed and separation ratios that are in quite good agreement with seismic data.  It is not possible to alter the sound speed in the solar core --- even at the 1$\sigma$ level illustrated by blue shading in Figure\,\ref{fig:sound} --- without simultaneously degrading the agreement with the separation ratios shown in Figure\,\ref{fig:ratios}. This suggests that the estimated uncertainty in the inversion procedure, while probably appropriate in 1997, is now in fact an overestimation of the true uncertainty. Note that separation ratios were introduced only later, and have benefited from determination of frequencies for low-angular degree modes from very long time-series data not available in the 1990s \citep{basu:2009}.

\subsection{Depth of convective zone and surface helium abundance}

The depth of the convective zone ($R_\mathrm{CZ}$) and surface helium fraction ($Y_\mathrm{S}$) are two traditional helioseismic constraints with which AGSS09met SSM also disagree. Helioseismic values and results from our SSM calculations are presented in Table~\ref{tab:seismic}. As before, we assume the same model uncertainties regardless of the solar composition, in order to allow a direct comparison between solar models. V16 have considered these observables using the linear solar models and their changes in central values are consistent with those computed from SSMs. As can be seen in Table~\ref{tab:seismic}, while $R_\mathrm{CZ}$  improves, the resulting $Y_\mathrm{S}$ is in serious conflict with the helioseismically-inferred value. The vSZ16 model leads to more than a 6$\sigma$ discrepancy, up from a 3.6$\sigma$ problem for AGSS09met. V16 claim an agreement at the 1.3$\sigma$ level, but this is only because their quoted error bars are very large. Under the same considerations, AGSS09met would be well within 1$\sigma$ of the helioseismic value.

The formal agreement claimed by V16 is in fact worse than that of AGSS09 when comparable abundance uncertainties are used for both cases. Only by virtue of using large error bars and applying them only to the model based on the vSZ16 composition, do V16 make the formal agreement seem better for $Y_\mathrm{S}$.

The question therefore arises: why is $Y_\mathrm{S}$ so much worse if the sound speed profile (although see Sect.~\ref{sec:combined} for further discussion on this issue) and $R_\mathrm{CZ}$ are both better in the vSZ16 model than in AGSS09met? The simple reason is that the initial helium of the model is much more sensitive to refractories than to volatiles, which is subsequently reflected in $Y_\mathrm{S}$. This is summarised in Table\,1 in \citet{serenelli:2010}, where power-law dependences of helium on different solar model input parameters are given. On the other hand, the sound speed profile and $R_\mathrm{CZ}$ are much more sensitive to the opacity profile around the base of the convective zone, where volatiles, in particular oxygen, play a dominant role. As the volatile abundances in vSZ16 are very close to those in GS98, agreement in these observables should not come as a surprise; the much more abundant refractories, however, lead to a strongly excluded surface helium abundance.

\begin{table}
\centering
\caption{Convective zone radius $R_\mathrm{CZ}$ and surface helium abundance $Y_\mathrm{S}$ for the solar models considered here, along with values inferred from helioseismology. 
Solar $R_\mathrm{CZ}$ is from \citet{basu:1997} and $Y_\mathrm{S}$ from \citet{basu:2004}. \label{tab:seismic}}
\begin{tabular}{lcc}
\hline
Model/Sun & $R_\mathrm{CZ}$  & $Y_\mathrm{S}$ \\ \hline
GS98 & $0.712\pm0.002$ & $0.243 \pm 0.003$ \\
AGSS09met & $0.723\pm0.002$ & $0.232 \pm 0.003$ \\
vSZ16 & $0.715\pm0.002$ & $0.277 \pm 0.003$ \\
Sun & $0.713\pm0.001$ & $0.2485\pm0.0034$ \\ \hline
\end{tabular}
\end{table}

\section{Combined analysis} \label{sec:combined}

Here we present a more rigorous quantitative analysis of the overall agreement between each of the three solar models and available limits, using $\Phi(^8{\rm B})$ and $\Phi(^7{\rm Be})$ solar neutrinos as well as helioseismic diagnostics. We follow the method presented in \citet{villante:2014}, which accounts for model correlations among different observables. In order to base our conclusions on the same helioseismic observables considered in V16, we do not include in this analysis the frequency separation ratios. Unlike the analysis in \citet{villante:2014}, because we want to test the different solar compositions, we fix the elemental abundances. All non-compositional input parameters (e.g. nuclear cross sections, microscopic diffusion rates, solar age) are allowed to deviate from their central values by introducing the so-called pulls and a penalty function to the $\chi^2$ calculation (see \citealt{villante:2014} for details about the statistical approach). Note that for this reason, when a combination of observables is considered, the resulting $\chi^2$ is not simply the addition of the individual contributions of each observable. 

Table~\ref{tab:chi2} presents the results for different sets of observables, taken into account one at a time, as labelled. The row $\{c_i \}$ corresponds to the sound speed profile, for which we consider 30 points distributed across the radiative interior \citep{basu:2009}. The final row in Table~\ref{tab:chi2} corresponds to all observables considered simultaneously.  The vSZ16 model is better than AGSS09met only for $R_\mathrm{CZ}$. For all other observables, the performance of the vSZ16 model is worse than AGSS09, in some cases by a large amount.

\begin{table}
\centering
\caption{Goodness of fit of each observable considered here, for each solar model.  \label{tab:chi2}}
\begin{tabular}{lccc}
\hline
 & $\chi^2_{\rm GS98}$  & $\chi^2_{\rm AGSS09met}$ & $\chi^2_{\rm vSZ16}$ \\ \hline
$Y_\mathrm{S}$ & 1.4 & 13.5 & 34.2 \\
$R_\mathrm{CZ}$ & 0.15 & 14.8 & 0.60 \\
$Y_\mathrm{S} + R_\mathrm{CZ}$ & 1.6 & 64.8 & 47.3 \\
$\{c_i \}$ & 46.4 & 111.2 & 359.3\vspace{0.5mm}\\
\hline
$\Phi(^8{\rm B})$ & 0.44 & 1.18 & 19.0 \\
$\Phi(^7{\rm Be})$ & 0.28 & 0.45 & 15.0\vspace{0.5mm}\\
\hline
Combined (34 dof) & 65.5 & 186.1 & 489.1 \\
\hline
\end{tabular}
\end{table}

The sound speed results deserve some comment. Looking at Fig.\,\ref{fig:sound}, it might seem that the vSZ16 model is closer to the Sun than AGSS09met.  In Table~\ref{tab:chi2} however, the resulting $\chi^2$ is actually worse for vSZ16. The reason is that variations in non-compositional input parameters, in particular an increase in the diffusion rate, lead to improvements in the sound speed profile of an SSM based on the AGSS09 composition in the region between 0.4 and 0.7\,${\rm \rsun}$, and thus help to partly reconcile its prediction with observations. On the other hand, for the vSZ16 model this is not possible because varying non-compositional parameters cannot improve the agreement in the region of strongest discrepancy, below 0.4\,${\rm \rsun}$, brought about by the large abundance of refractories. 

The conflicts with observation that arise in an SSM based on the vSZ16 composition are in fact worse than the problem this solar model was supposed to cure, i.e. the discrepancy between helioseismic data and SSMs based on low-Z solar compositions such as that from AGSS09.  This alone should be a good indication that the vSZ16 composition is unlikely to be representative of that in the solar interior. The next section gives even more direct reasons why the vSZ16 abundances cannot be representative of the composition of the solar photosphere.

\section{Abundances from the solar wind} \label{sec:compo}

The results and essentially all conclusions of \citet{v16} rest on the correctness of a quite non-standard set of solar abundances, originating from solar wind measurements  \cite{vS10,vS16}.  
These are based on \textit{in situ} analysis of ions in the wind emerging from polar coronal holes (PCHs).  Solar activity predominantly affects the corona and wind at low latitudes, leaving both the rate and composition of the wind emerging from polar regions approximately constant over time.  The wind from PCHs is therefore understood to be indicative of the underlying steady state of mass emission from the Sun.  \cite{vS10} and \cite{vS16} claim that this makes the wind from PCHs the least affected by fractionation effects, which are known to impact the relative abundances of different nuclei in other solar wind samples.  This supposedly allows the derivation of photospheric abundances from samples of the solar wind originating in PCHs. In fact, using these abundances to construct SSMs implicitly assumes they match photospheric values, i.e. that there is no fractionation at all. We show below that this assumption is incorrect.

\subsection{Spectroscopy and astrophysics}

Before looking at the solar wind measurements themselves, it is worth thinking about the basic plausibility of the solar composition advocated by \cite{vS16}, from the spectroscopic and astrophysical perspectives. The PCH-based CNO abundances in Table \ref{tab:compo} are indeed uniformly higher than AGSS09, but in rough agreement with GS98 values.  The abundances of refractory elements (Mg, Si, S and Fe) are however a full 0.3\,dex higher in general than AGSS09.  This makes them far higher than any spectroscopically-determined abundances in over half a century, including the pioneering works of \citet{Goldberg60}, \citet{RossAller} and \citet{AG89} -- let alone the solar abundances presented by GS98, AGSS09, \citet{Lodders09} or \citet{caffau:2011}.  This includes determinations based on 1D and 3D model atmospheres, with and without corrections for departures from local thermodynamic equilibrium (LTE), and with old and new atomic data.  Taken as presented by \citet{vS16}, the abundances of refractory elements cannot be reconciled with the results of any spectroscopic determination, regardless of its sophistication.

For spectroscopic analysis to be this mistaken on the refractory abundances, one or both of the following exceptionally unlikely scenarios would have to be true:\begin{enumerate}
\item All oscillator strengths measured by dedicated atomic physics laboratories around the world for Si, S and Fe are systematically overestimated by about a factor of 2, as are the theoretical values computed for Mg.  Each experimental atomic physics group does its work independently, employing sophisticated and accurate modern techniques like laser-induced fluorescence for determining the absolute scales of their transition probabilities, and cross-checks the results with entirely different techniques.  This produces errors better than 5\% in many cases.
\item The basic underlying theoretical or methodological framework of stellar atmospheres is somehow wrong, due to something systematically amiss in all calculations of radiative transfer or atmospheric modelling --- not merely in the specific application of 3D atmospheric models and line formation modelling.  This would invalidate the entire field of stellar atmospheres, and stellar abundance analysis generally. This would require discarding an enormous number of bedrock astrophysical results, with wide-ranging and highly implausible implications for the mutual consistency of stellar nucleosynthesis, stellar evolution, Galactic chemical evolution and even cosmology.
\end{enumerate}

It is also worth remembering that the AGSS09 composition is  consistent with the Sun being an otherwise unremarkable Galactic thin-disk G dwarf, showing good agreement with expected abundance patterns in the nearby neighbourhood.  These range from measurements of abundances in so-called `solar twins' \citep{Melendez09,Ramirez09}, to comparisons with young B-type stars \citep{Nieva11,Nieva12}, local H\,\textsc{ii} regions \citep{Esteban04,Esteban05} and the local interstellar medium \citep{Henry10}.  In particular, the local `cosmic abundance standard' $Z=0.014\pm0.002$ \citep{Nieva12}, is in agreement with the AGSS09 solar metallicity $Z=0.0134$, and incompatible with the \citeauthor{vS16} value of $Z=0.0196$. Note that B-type stars have radiative atmospheres, so systematic uncertainties that might be associated with near-surface convection in the solar atmosphere do not play any role. 

\subsection{Solar wind}
\label{abundances}

Spectroscopic and astrophysical considerations strongly suggest that the composition presented by \citet{vS16} should not be trusted as representative of the photosphere or the bulk Sun.  Where then is the neglected systematic error (or errors) in the solar wind analysis?  There appear to be two distinct but related sources.  The first is apparent fractionation in the PCH sample relative to the photosphere, and the second is the normalisation scale and associated uncertainties used by \cite{vS16} to compare their abundances to the photosphere.

\begin{figure}
\includegraphics[width=\columnwidth]{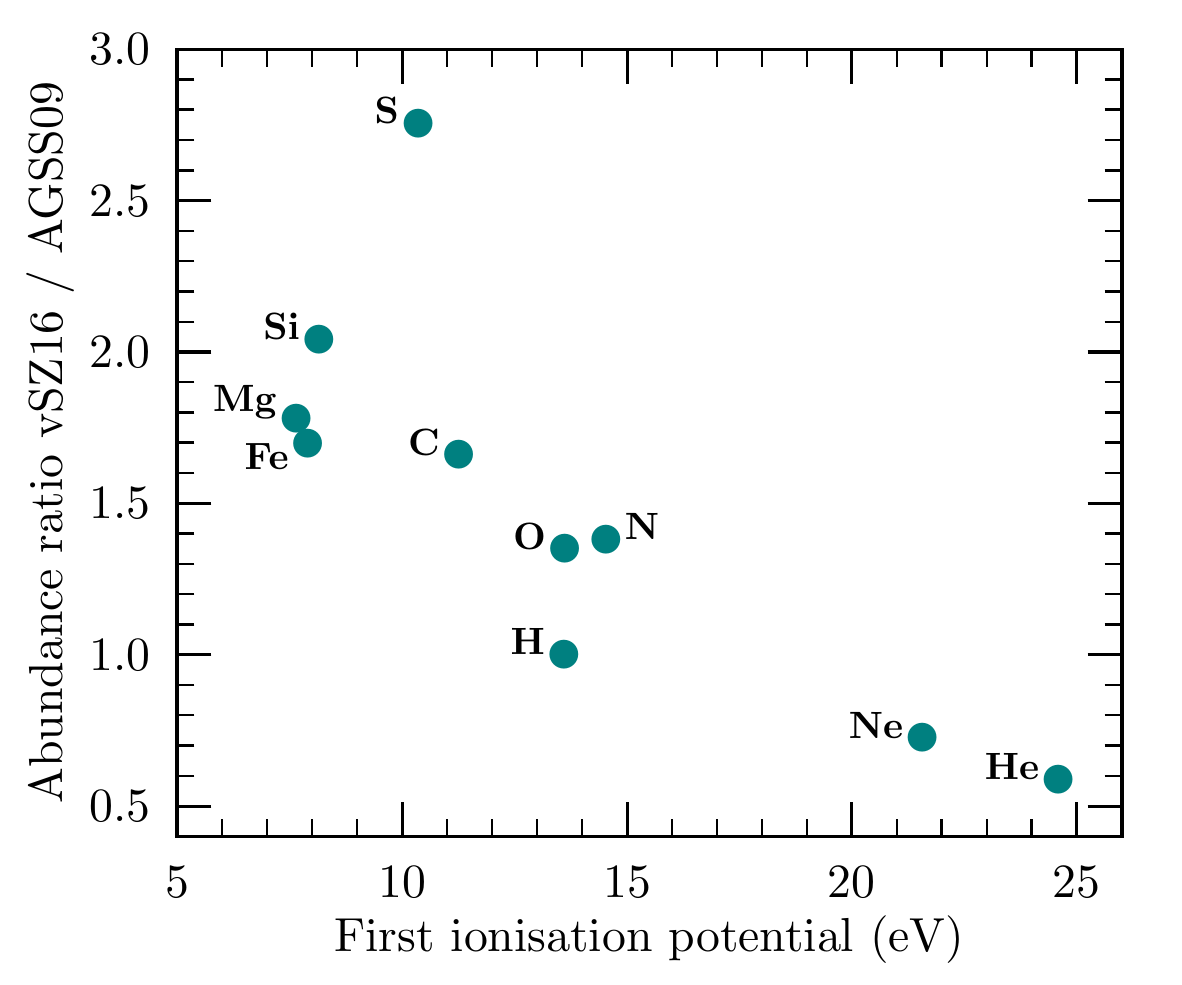}
\caption{The ratio of solar abundances advocated by \citet[][vSZ16]{vS16} and \citet[][AGSS09]{agss09}.  The trend with first ionisation potential indicates that despite being based on wind originating from polar coronal holes, the vSZ16 abudances are affected by fractionation arising from the FIP effect. \label{fig:abun}}
\end{figure}

The first ionisation potential (FIP) effect is well known to affect the relative abundances of elements in the solar wind, enhancing the abundances of elements with low ionisation potentials, and reducing those with higher ionisation thresholds \citep{Laming15}.  This effect is thought to be reduced in PCH outflows compared to equatorial winds, in part because the composition is known to vary with activity in the latter \citep{Z02}.  However, the stability of the composition of the solar wind from PCHs is not in and of itself an indication that the wind is unfractionated.  Indeed, Fig.\,\ref{fig:abun} shows that the abundance discrepancy between AGSS09 and vSZ16 exhibits a strong trend with the first ionisation potential of the elements considered.  This is clear indication that the PCH abundances are still fractionated, even if less so than other solar wind samples --- a problem also noticed by \citet{Laming15}.

It is also notable from Fig.\,\ref{fig:abun} that the FIP effect can \textit{increase or decrease} the abundance of an element, depending on whether it possesses a first ionisation potential greater or smaller than the reference element used for setting the abundance scale.  In this case that is hydrogen, so helium and neon are depleted relative to the true photospheric values, whereas other elements are enhanced.  This explains the implausibly high refractory abundances of vSZ16, and falsifies their claim that their value of $Z$ is a lower bound because unquantified fractionation would only decrease $Z$.

Indeed, H is not the logical reference element to choose when compiling abundances from the solar wind.  All the abundances of vSZ16 are based on measurements of elemental ratios with respect to O, set to the usual spectroscopic hydrogen scale using a single measurement of H/O = $1500\pm300$ (corresponding to $\log \epsilon_{\rm O} =8.82^{+0.10}_{-0.08}$) by \citet{vS10}.\footnote{We note that the more recent analysis of the solar wind composition by \citet{2013ApJ...768...94L} instead found a most probable value of $\log \epsilon_{\rm O} =8.68$ for the fast solar wind.}  \citet{vS16} neglected to include the systematic uncertainty of the H/O normalisation in their adopted abundances, drastically reducing the error budget in comparison to the correct calculation.  Propagating the error from the normalisation and combining it in quadrature with the errors on the individual X/O ratios, the uncertainties on the abundances of vSZ16 can be seen to typically exceed 0.1\,dex, as shown in Table\ \ref{tab:compo}.  For CNO, the coarse abundances obtainable from the solar wind are in fact consistent with the more precise values in AGSS09.  This is in large part due to the similarity of the ionisation potentials of H, C, N and O; the erroneous nature of the vSZ16 refractory abundances persists.  It is surprising that \cite{vS16} failed to include this important systematic uncertainty, yet somehow saw fit to claim that AGSS09 did not include systematic errors --- despite the fact that careful quantification and inclusion of systematic errors from non-LTE, the mean temperature structure of the adopted models, and the impact of 3D effects, was one of the key advances highlighted in AGSS09.

A less error-prone way to present solar wind abundances would be to choose O as the common element of comparison, removing any systematic uncertainty due to the absolute scale, in a similar way that Si is chosen for comparison with CI chondritic meteorites.  This would of course also substantially reduce the central value of the overall metallicity implied by the measurements of vSZ16, and completely change the resulting solar models of \citet{v16}.  Indeed, given the trend in Fig.\ \ref{fig:abun}, there is no good reason to think that the H/O ratio of \citet{vS10} is free of additional unquantified fractionation effects anyway.  It is quite possible that even for elements with common first ionisation potentials, some additional effect (sub-dominant to the FIP but visible nonetheless) is causing fractionation at a level beyond the uncertainty in the photospheric abundances.  This is unsurprising really, given that in spite of much theoretical work the FIP is still poorly understood -- especially its quantitative impact on elemental abundances. Higher ionisation potentials must surely play a role as well, given that the \textit{in situ} measurements involve higher charged states of each species.

\section{Summary}
We have shown that solar models constructed from the chemical composition advocated by \citet{vS16} and \citet{v16} provide vastly worse fits to the observed neutrino fluxes, sound speed profile and surface helium fraction of the Sun compared to those constructed from the canonical AGSS09 mixture that gave rise to the solar modelling problem; only the radius of the convective zone is improved.  We have also demonstrated that the composition of \citet{vS16} is subject to large, unquantified normalisation and fractionation errors, and can be safely ruled out on spectroscopic and astrophysical grounds.  The solar modelling problem persists: accommodating helioseismology data with the best determined solar abundances and the best standard solar models is still an unsolved problem.

\section*{Acknowledgements}

A.S. is partially supported by grants ESP2014-56003-R and ESP2015-66134-R (MINECO) and 2014-SGR-1458 (Generalitat de Catalunya). P.S. is supported by STFC (ST/K00414X/1 and ST/N000838/1). The work of  F.L.V. is supported by the Italian Ministero dell'Istruzione, Universit\`a e Ricerca (MIUR) and Istituto Nazionale di Fisica Nucleare (INFN) through the ``Theoretical Astroparticle Physics'' research projects. S.B. acknowledges partial support from NSF grant AST-1514676 and NASA grant NNX13AE70G. C.P.G. is supported by Generalitat Valencia Prometeo Grant II/2014/050, by the Spanish Grant FPA2014-57816-P of MINECO and by PITN- GA-2011-289442-INVISIBLES.

%%%%%%%%%%%%%%%%%%%%%%%%%%%%%%%%%%%%%%%%%%%%%%%%%%

%%%%%%%%%%%%%%%%%%%% REFERENCES %%%%%%%%%%%%%%%%%%

% The best way to enter references is to use BibTeX:

%\bibliographystyle{mnras}
%\bibliography{example} % if your bibtex file is called example.bib

% Alternatively you could enter them by hand, like this:
% This method is tedious and prone to error if you have lots of references

% Don't change these lines
\bsp	% typesetting comment
\label{lastpage}
\end{document}